\documentclass[12pt]{article}

\usepackage{amstext,amsmath,amssymb,amsfonts,bbm}
\usepackage[latin1]{inputenc}
\usepackage{epsfig}
\usepackage{hyperref}
\usepackage{amsthm}
\usepackage{subfigure}
\usepackage{color}
\usepackage{multirow}
\newtheorem{lemma}{Lemma}

\def\beq{\begin{equation}}
\def\be{\begin{equation}}
\def\ee{\end{equation}}
\newcommand{\bea}{\begin{eqnarray}}
\newcommand{\eea}{\end{eqnarray}}

\newcommand{\unit}{\mathbbm{1}}

\def\f{\frac}
\def\mone{^{-1}}

\def\C{{\mathbbm C}}
\def\N{{\mathbbm N}}
\def\Z{{\mathbbm Z}}
\def\R{{\mathbbm R}}

\DeclareMathOperator{\Tr}{Tr}

\DeclareMathOperator{\Hom}{Hom}
\DeclareMathOperator{\Hess}{Hess}
\DeclareMathOperator{\SU}{SU}
\DeclareMathOperator{\vol}{vol}

\DeclareMathOperator{\ext}{ext}
\DeclareMathOperator{\inte}{int}

\newcommand{\alg}{\mathfrak{g}}
\newcommand{\su}{\mathfrak{su}}

\newcommand{\cG}{{\mathcal G}}

\makeatletter

\begin{document}

\begin{titlepage}
\begin{flushright}
pi-qg-208\\
ICMPA-MPA/002/2011
\end{flushright}

\vspace{20pt}

\begin{center}

{\Large\bf Radiative corrections in the Boulatov-Ooguri \\
\vspace{10pt}
tensor model: The 2-point function}
\vspace{20pt}

Joseph Ben Geloun$^{a,b,*}$, and Valentin Bonzom $^{a,\dag}$

\vspace{15pt}

$^{a}${\sl Perimeter Institute for Theoretical Physics}\\
{\sl 31 Caroline St. N., ON, N2L 2Y5, Waterloo, Canada}\\
\vspace{10pt}
$^{b}${\sl International Chair in Mathematical Physics and Applications\\ (ICMPA-UNESCO Chair),
University of Abomey-Calavi,\\
072B.P.50, Cotonou, Rep. of Benin}\\

\vspace{20pt}

E-mail:  $^{*}${\em jbengeloun@perimeterinstitute.ca},\quad 
$^{\dag}${\em vbonzom@perimeterinstitute.ca}

\vspace{10pt}

\begin{abstract}
\noindent
The Boulatov-Ooguri tensor model generates a sum over spacetime topologies for the $D$-dimensional BF theory. We study here the quantum corrections to the propagator of the theory. In particular, we find that the radiative corrections at the second order in the coupling constant yield a mass renormalization. They also exhibit a divergence which cannot be balanced with a counter-term in the initial action, and which usually corresponds to the wave-function renormalization.

\end{abstract}
\end{center}

\noindent  Pacs numbers:  11.10.Gh, 04.60.-m, 04.60.Pp

\end{titlepage}

\setcounter{footnote}{0}

\section{Introduction}
\label{sect:intro}

The setting of group field theory (GFT) produces tensor models which generalize matrix models to structures of dimensions higher than two. In particular, the Boulatov model \cite{boul} generates a sum over topologies of 3-manifolds (as well as other objects) weighted by the corresponding Ponzano-Regge amplitude, which is a model for the partition function of 3d Riemannian quantum gravity. As a field theory, it provides a good framework to address the renormalization of theories where one has to sum over topologies of the spacetime manifold (or cell complex). In particular, it stands as a promising approach to quantum gravity, via spin foam models, where one aims at a background free quantization of general relativity, \cite{Freidel,oriti}-\cite{Oriti:2009wn}.

Several interesting results and advances in GFTs  have been obtained during the last years. From contacts with simplicial geometry \cite{oriti3} to mechanisms for emergent matter fields (incorporating a new/extra gauge symmetry) \cite{DiMare:2010zp}, from contacts to noncommutative geometry \cite{Baratin:2010wi} and quantum groups \cite{Girelli:2010ct,Baratin:2011tg} to classical geometrodynamics extraction and Bose condensate techniques \cite{Oriti:2010hg}, indeed GFT evolves fast and in several directions.

From the quantum field theory perspective, efforts have been concentrated so far mostly on power-counting results, either in the Boulatov model and its higher dimensional extensions (coined flat spin foam model) \cite{fgo}-\cite{Bonzom:2010ar}, or in theories hopefully related to quantum gravity, \cite{Krajewski:2010yq,Geloun:2010vj}. An exciting development of these power-counting results is the recently defined  large $N$ limit, \cite{Gurau:2010ba}, for colored GFTs \cite{gurau}-\cite{Caravelli:2010nh} 
where the analog of the planar graphs of matrix models 
are found to be 2-complexes corresponding to the 3-sphere.

Still, one may be interested in making sense of the theory without the $1/N$ expansion. A key feature which remains to be unraveled is a locality principle -- typically in ordinary scalar field theory, it says that renormalization appears when probing high internal scales, i.e. short distances, with low external momenta. For GFTs, a generalized locality principle was seen to emerge in \cite{Krajewski:2010yq}. Also in \cite{Bonzom:2010zh}, the flat spin foam model was considered as a generalization of the zero coupling limit of 2d Yang-Mills theory, \cite{2d-YM}, to cell 2-complexes. Thus, the set of interest which was used to localize the integrals was the set of flat connections on the complex, and the divergence degree could be extracted in terms of a discrete analog of the twisted De Rham cohomology for the covariant differential $d_A$, $A$ being a flat connection.

In this paper, we will use the setting which proved fruitful in \cite{Bonzom:2010zh} to compute radiative corrections to the propagator. It is worth emphasizing the main observation which makes this method efficient. The manifold in the model is (several copies of) a compact Lie group, say $\SU(2)$. Thus, the momentum space is described in terms of half-integer spins (irreducible representations of $\SU(2)$). To understand the localization in the Boulatov model, it is much simpler to explicitly perform the sums over the spins, which produce delta distributions on the manifold. Getting a similar picture of the localization without the explicit result of summing over the spins is certainly a challenging task, and it may help to extend the analysis to other models (where the integrals may be concentrated around something else than flat connections).

To make our program on the 2-point function clear, let us remember the kind of expansion usually performed in the ordinary $\phi^4_4$ theory, and which leads to the key ideas of mass subtraction and wave function renormalization. The standard textbook procedure is to re-express the bare quantities in terms of the physical ones, say the field: $\phi_b = Z^{1/2}\phi_r$, and the mass $m_r^2  = m_b^2 Z - \delta_m$, so that the Lagrangian, written with renormalized quantities, picks up counter-terms of the form:
\beq
p^2\,\delta_Z - \delta_m\;,
\ee
in momentum space, with $\delta_Z = Z-1$. Obviously, the term proportional to $p^2$ comes from the part of the action with derivatives of the field, $\phi\, \partial^2 \phi$ in direct space. Then, $\delta_m, \delta_Z$ are extracted from radiative corrections to the 2-point function, using some renormalization prescription. This is done by evaluating the graph amplitude at a given external scale, like $p^2_{\rm ext} = m^2_r$, for $\delta_m$, and evaluating its derivatives with respect to $p^2$ at this scale for $\delta_Z$.

Note that in \cite{Geloun:2010vj} the mass subtraction has been roughly worked out for the EPRL/FK model, a candidate for quantum gravity via spin foams. Nevertheless, we would like to understand first how the prime concepts of leading divergences and counterterms find themselves a natural extension in the simpler Boulatov-Ooguri tensor model.

This model differs from the usual $\phi^4_4$ theory by the fact that the quadratic part of the Lagrangian does not contain derivatives of the field, but only a mass term,
\beq \label{rough action}
\int m^2\ \phi^2 + \lambda\ \phi^4\;.
\ee
So, roughly\footnote{The peculiarities of the GFT (being non-local, the covariance projecting the field onto a rotationally invariant sector) play no role in this discussion.}, the propagator behaves like: $C = \int dp \frac{e^{ipx}}{m^2}$. Consequently, we will look for organizing the expansion of each 2-point graph, not as an asymptotic expansion with the cut-off, but as a Taylor expansion where the $n$-th order is proportional to a $n$-th derivative $C^{(n)}$ of the propagator. Symbolically:
\beq \label{typical exp}
 \Lambda^{k_0}\,C   +  \Lambda^{k_1}\,C^{(1)}  + \Lambda^{k_2}\,C^{(2)}  + \dotsc
\;,
\ee
where $\Lambda$ is the ultraviolet cut-off. We are interested only in the leading orders of this expansion, i.e. the terms which diverge when the cut-off $\Lambda$ goes to infinity. A divergence with the cut-off at the zeroth order, i.e. $k_0\geq 0$, is a constant term in momentum space (given the specific form of the propagator), and gives a mass renormalization $\delta_m$. We expect that there is no term with $C^{(1)}$ by symmetry. The term proportional to $C^{(2)}$ behaves like $p^2$ in momentum space. Thus if the corresponding leading term diverges with $\Lambda$ (i.e. $k_2\geq 0$), we should get a wave-function renormalization $\delta_Z$.

This is indeed what we will find, from a graph at the second order in the coupling constant. However, due to the form of the initial action \eqref{rough action}, there is no possible counter-terms for the divergences with $C^{(2)}$. This is the sign that the action needs an addition of a quadratic term with second derivatives of the field, so that it takes the form:
\beq
\int \phi\,\Delta\phi + m^2\, \phi^2 + \lambda\, \phi^4\;,
\ee
where $\Delta$ is the Laplace operator on the group manifold. It turns out that the resulting group field theory has already been considered in the literature by Oriti (\cite{DiMare:2010zp} and more references therein) with different motivations.

The organization of the paper is as follows. In Section \ref{sec:action}, we define the Boulatov -Ooguri tensor model.
\begin{itemize}
 \item In Section \ref{sect:1}, we study a correction to the 2-point function, at the second order in the coupling constant, and observe two leading divergences. We show that there is indeed a mass renormalization. We also find a divergence proportional to the second derivatives of the propagator, which cannot be balanced with a counter-term from the initial action. The structure of the corresponding term is studied with details in the 3d case.
 \item In Section \ref{sec:gf}, we study generic graphs of the 2-point function. We obtain an expression, in Lemma \ref{lemma:gf}, which shows that the integrand is localized on the set of flat connections on the 2-complex, so that the generalized Laplace approximation of \cite{Bonzom:2010zh} applies. The first order in particular is an averaging of the propagator with non-trivial insertions on flat connections.
 \item In Section \ref{sec:simplyconnected}, we show that an expansion of the form \eqref{typical exp} is directly available for a class of 2-point graphs which are simply connected 2-complexes. We then find that there exist divergences to the order $n$ in the derivatives of the propagator with $n$ arbitrarily high, which are computed from Gaussian moments of the zero order.
 \item Section \ref{concl} discusses the addition of specific quadratic terms with second derivatives of the field in the action. We extract the propagator, and argue that we have a natural setting for a scale analysis.
\end{itemize}

The technical tools we use do not rely on the specific form of the graphs of the GFT, and may be applied to some cellular complexes as well. Nevertheless, there is no hope to get an expansion like \eqref{typical exp} for the generic GFT graphs. The kind of graphs we will consider throughout the paper contains all graphs of the colored model \cite{gurau}, a subset of the generic graphs, and certainly graphs which are not of the GFT type.

\section{The action and the graph amplitudes} \label{sec:action}

We start by giving the action and some properties of Feynman graphs for the $D$-dimensional GFT of Boulatov-Ooguri type, \cite{boul}, over a compact Lie group $G$. Typically, $G = SU(2)$ is considered throughout the text. Then in the subsequent sections, we proceed to a specific expansion of the 2-point function.

Fields belong to the Hilbert space of square integrable functions on $G^{D}$, namely $H= L^2 (G^{D})$. These fields are actually restricted the so-called {\em gauge invariant} fields, in the sense that they are invariant under the diagonal right action of the group:
\begin{equation}
\phi(g_1h,g_2h,\dotsc, g_D h)=\phi(g_1,g_2,\dotsc,g_D)\;, \quad \forall h \in 
G\;. 
\label{inv}
\end{equation}
The quadratic part of the action has the form: 
\begin{equation}
\int\prod_{s=1}^{D}dg_s\ \phi(g_1,\dotsc,g_D)\phi(g_1,\dotsc,g_D)\;,
\end{equation}
where $dg$ denotes the Haar measure on $G$. Due to the restriction \eqref{inv} to a subset of fields, the propagator is not just the inverse of this quadratic part. Instead, one has to use a (normalized) degenerate Gaussian measure $d\mu_C[\phi]$, of covariance $C$ defined by:
\begin{equation} \label{propa}
C(\{g_1,\dotsc,g_D\}; \{\tilde{g}_1,\dotsc,\tilde{g}_D\}):= \int dh\ \prod_{s=1}^D 
\delta(g_{s}h\tilde{g}_{s}^{-1})\;.
\end{equation}
The integral on $h$ is a group averaging which enforces the condition \eqref{inv}.

The interaction is non-local, of degree $(D+1)$, namely a $\phi^{D+1}$ theory:
\bea
S_{int}[\phi]&:=&
\lambda \int \prod dg_{a^b}\  
\phi_{1^2,1^3,\dots,1^{D+1}}\;
\phi_{(D+1)^1,(D+1)^2,\ldots,(D+1)^D}\;
\phi_{D^{D+1},D^1,\ldots,D^{D-1}}\dots \crcr
&&\phi_{3^4,3^5,\ldots,3^{D+1}, 3^2}\;
\phi_{2^3,2^4,\ldots,2^{D+1}, 2^1}
\prod_{a\neq b}^{D+1}\delta(g^{\phantom{-1}}_{a^b} g^{-1}_{b^a})\;,
\eea
where the shorthand notation $\phi(g_{l^a},g_{l^b},\dots)= \phi_{l^a,l^b,\dots}$ is understood. Each integration variable appears in two copies of the field.

The partition function is:
\begin{equation}
\label{formal}
Z(\lambda):=\int d\mu_C[\phi]\ e^{-\lambda S_{int}[\phi]}\;.
\end{equation}

\begin{figure}[htb]
\centering{
\includegraphics[width=85mm]{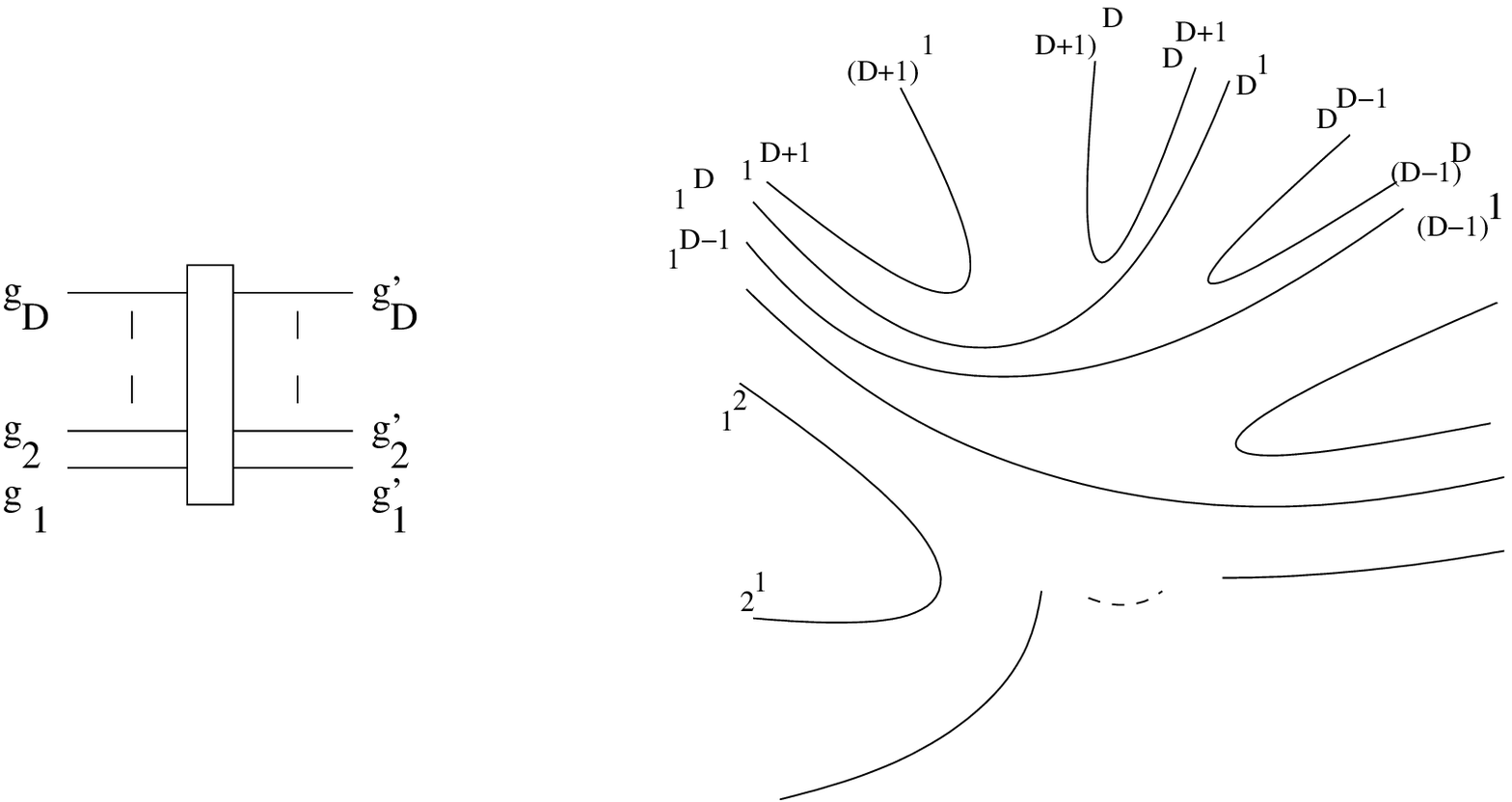}}
\caption{$D$ dimensional model propagator and vertex.}
\label{fig:prop2}
\end{figure}

The propagator can be drawn with $D$ strands (see Figure \ref{fig:prop2}), each of them corresponding to a delta function in \eqref{propa}. As for the vertex, each of the $(D+1)$ legs has $D$ strands. One strand goes between two legs, and there is a single strand between any pair of legs. The Feynman rules produce Feynman graphs which are actually 2-complexes. Indeed, in addition of the set of vertices and edges (or lines, due to propagators), one identifies faces as regions being bounded by closed strands. An open strand gives an open face. The corresponding amplitudes take the form:
\bea
&&
A_{\mathcal G}(\{g^{\ext}_s\};\{\tilde{g}^{\ext}_s\}) = \crcr
&&\crcr
&&
\int  \prod_{\ell\in L(\mathcal G)} dh_\ell  \left[
\prod_{f \in F^{\ext}(\mathcal{G})}
\delta \left(g^{\ext}_{s} \left[ \prod_{ \ell \in \partial f}  h^{\epsilon_{lf}}_\ell  \right] 
(\tilde{g}^{\ext}_{s})^{-1}\right) \right]
\left[ \prod_{f \in F^{\inte}(\mathcal{G})}
\delta \left( \prod_{ \ell \in \partial f}  h^{\epsilon_{lf}}_\ell  \right) \right],
\label{amp1}
\eea
where the group elements $\{g^{\ext}_s \}_{s=1}^D$ and $\{\tilde{g}^{\ext}_s\}_{s=1}^D$ are the arguments on the external legs, $ L(\mathcal G)$ denotes the set of (internal) edges or lines, $F^{\ext}(\mathcal G)$ and  $F^{\inte}(\mathcal G)$ the sets of external (open) and internal (closed) faces, respectively, of the graph $\mathcal G$. The product over ``$l \in \partial f$'' means the product over the set of lines $l$ belonging to the boundary of the face $f$. The sign $\epsilon_{lf}=\pm 1$ is equal to $+1$ if the orientations of the face and of the line coincide, to $-1$ if not and to $0$ if the line does not belong to the boundary of $f$. Note that we take, as a convention and without loss of generality, a particular orientation of the external leg arguments. 

There is a natural geometric interpretation which generalizes the two-dimensional interpretation of graphs of matrix models to higher-dimensional cell complexes. Indeed, the specific pattern of the vertex mimics the structure of a $D$-simplex. Each leg is considered to be one of the $(D+1)$ boundary $(D-1)$-simplices. A strand represents a $(D-2)$-simplex on the boundary, shared by exactly two $(D-1)$-simplices. In this view, the propagator is seen as a $(D-1)$-simplex. Furthermore, the Feynman rules produce all possible ways of gluing a set of $D$-simplices along their boundary.

Since products of distributions are not always well-defined, formulae such as (\ref{amp1}) suffer from what one should consider from the quantum field theoretic point of view as divergences. In the present situation, they will be referred to as ``ultraspin'' divergences.

\section{First radiative corrections to the 2-point function}
\label{sect:1}

\begin{figure}[htb]
\centering{
 \includegraphics[width=60mm]{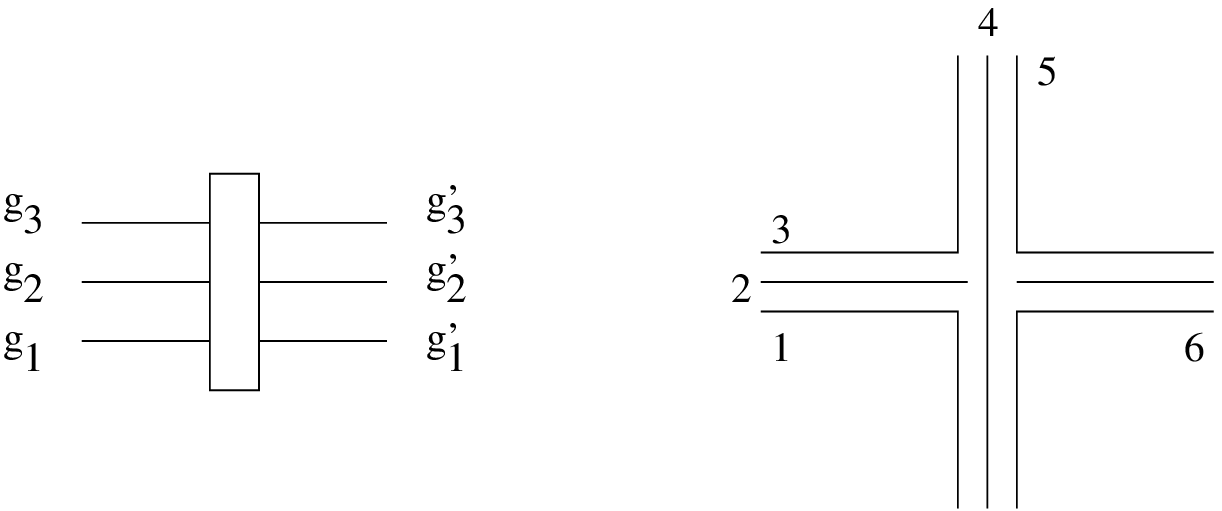}}
\caption{3d model propagator and vertex.}
\label{fig:prop}
\end{figure}

In three dimensions, a graph is built with 
three strands per propagator and the vertex is of the type $\phi^4$, see Figure \ref{fig:prop}.

\subsection{The first 2-point graph and its regularization} 

\begin{figure} 
\centering{
\includegraphics[width=60mm]{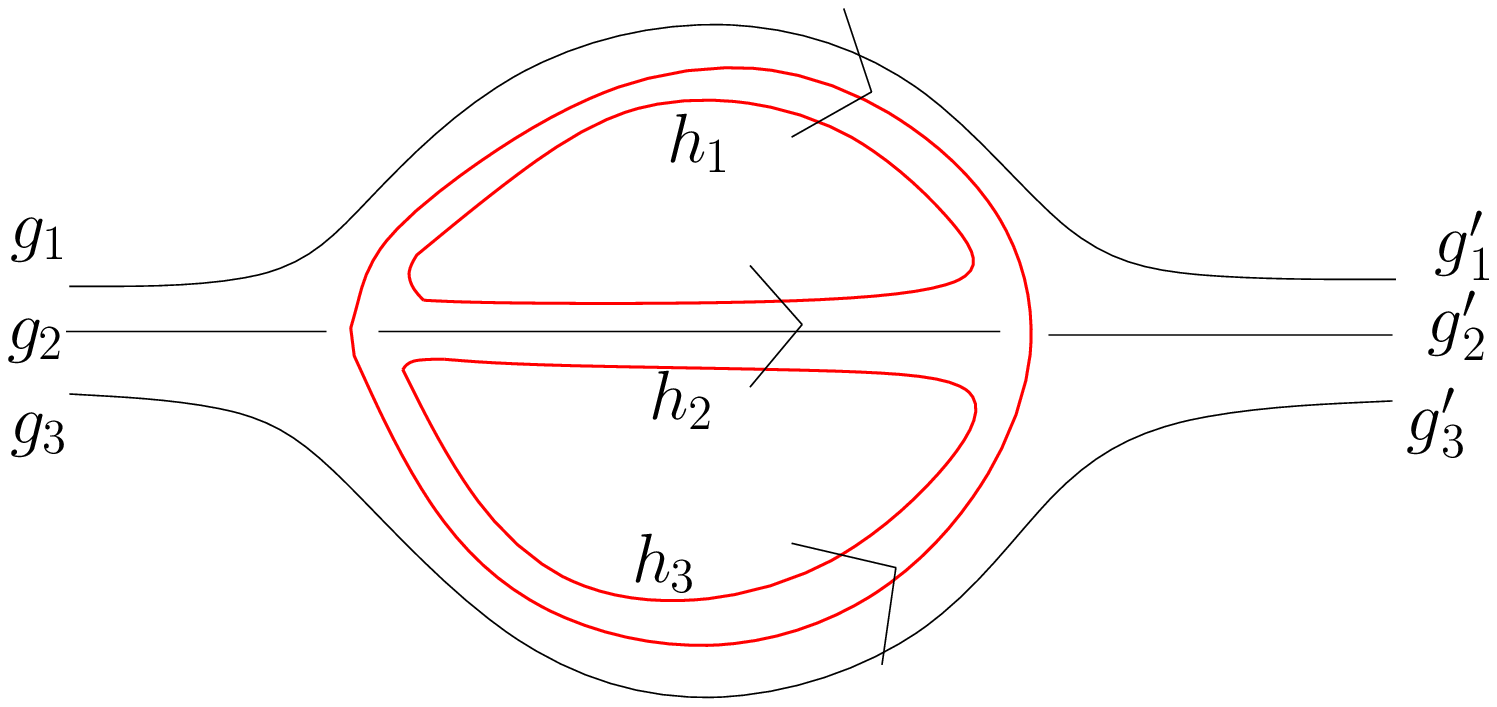}}
\caption{2-point function.}
\label{fig:2point}
\end{figure}

We begin our study with a 2-vertex graph of the 2-point function\footnote{In the ordinary GFT, there are lots of other 2-vertex graphs. But it is worth observing that the graph we consider is the simplest one in the {\em colored} GFT of \cite{gurau}.} (see Figure \ref{fig:2point}). The combinatorial ingredients of the graph are: three internal edges, 
$ L(\mathcal G)=\{l_1,l_2,l_3\}$, 
three external faces,
$ F^{\ext}(\mathcal{G}) = \{f^{0}_{1},f^{0}_{2},f^{0}_{3}\}$, 
and three internal faces,
$ F^{\inte}(\mathcal{G}) = \{f_{12},f_{13},f_{23}\}$.
Explicitly, the amplitude built from the graph formally reads:
\bea
A_{\mathcal G}(\{g_s\};\{\tilde{g}_s\}) = 
\int  \prod_{l=1}^3 dh_l\, 
\prod_{s=1}^3
\delta 
\left(g_{s} h_s (\tilde{g}_{s})^{-1}\right)
 \prod_{1\leq i <j \leq 3}
\delta 
\left( h_i h^{-1}_j  \right) \;.
\label{amp2}
\eea

To make sense of the product of Dirac delta in \eqref{amp2}, we change each of them with a heat kernel\footnote{The heat kernel is the solution of: $(\partial_t - \Delta)\,K_t(g)=0$, in which $\Delta$ is the Casimir-Laplace operator on $G$, and with the initial condition $\lim_{t\to 0}K_t(g) = \delta(g)$.} at time $t$ on $G$. Thanks to the Peter-Weyl theorem, it can be expanded onto $\SU(2)$ irreducible representations, labelled by half-integers (spins),
\beq
K_t(g) = \sum_{j \in \N/2} (2j+1)\,e^{-t C(j)}\ \chi_j(g)\;.
\ee
Here $C(j)$ is the Casimir in the representation of spin $j$, and $\chi_j$ the character. 
When $t$ goes to zero, this goes to $\sum_j (2j+1)\chi_j(g)=\delta(g)$, 
which is indeed the usual expansion of $\delta(g)$.

For small times $t$, the heat kernel is localized around zero, 
and its behaviour is close to that of the Euclidean kernel. 
In a neighborhood of the identity,
\be \label{asymp heat}
K_t(g) \underset{t\rightarrow 0}{\sim} \Lambda_{t}^{\dim G}\ e^{-\f{|g|^2}{4t}}\;,
\ee
where $\lvert g\rvert$ is the Riemannian distance from the identity to $g$, and 
\be
\Lambda_{t}\equiv(4\pi t)^{-1/2}\;.
\ee 

Results concerning the divergence degree of graphs in this theory 
can be found in the literature either using this heat kernel regularization, 
or using a sharp cut-off $\Lambda$ on the spin expansion
of the delta function, $\delta_\Lambda(g)= \sum_{j=0}^{\Lambda}
(2j+1) \chi_{j}(g)$. The cut-off $\Lambda_t$ is chosen 
so that the divergence degree of a given graph is the same if computed using $\delta_\Lambda$. 
Hence, in a nutshell, high spins correspond to small $t$.

As we explain in the paragraph below, we will apply a saddle point approximation to evaluate the graphs. However, we will be interested only in some specific terms which diverge with the cut-off $\Lambda_t$. For our purpose the asymptotic behaviour \eqref{asymp heat} will be sufficient. However to perform the full asymptotical expansion\footnote{ \label{foot}From a field theory perspective, such a full asymptotical series is not really relevant, since it will strongly depend on the chosen regularization. The most natural thing to do instead is to go within a multiscale analysis, in which one gets interested into subgraphs carrying higher scales than the rest of the graph.} in exponents of $\Lambda_t$, one would have to look at higher orders in the expansion of the heat kernel.

\subsection{Extracting the divergences: Mass and wave-function renormalizations} 
\label{sec:ex saddle}

The regularized amplitude can be expressed as
\beq
A_{\mathcal G; t}(\{g_s\};\{\tilde{g}_s\}) = \int  dh_1 dh_2 dh_3\,
\left[\prod_{s=1}^3
\delta \left(g_{s} h_s (\tilde{g}_{s})^{-1}\right)\right]
K_t \left( h_1\,h_2\mone \right) K_t \left(h_1\, h_3\mone\right)
K_t \left( h_2\,h_3^{-1}\right) \;.
\label{amp4}
\ee
Changing of variables $ k_i = h_1^{-1} h_i$, for $i=2,3$, and using the translation invariance of the Haar measure allow us to rewrite the same amplitude as
\begin{multline}
A_{\mathcal G;t}(\{g_s\};\{\tilde{g}_s\}) =
\int dk_2 dk_3\,\left[\int dh_1
\delta  \left(g_{1} h_1 (\tilde{g}_{1})^{-1}\right)
\prod_{s=2}^3
\delta 
\left(g_{s} h_1k_s (\tilde{g}_{s})^{-1}\right)
\right]\\
K_t( k_2 )K_t( k_3)
K_t( k_2k_3^{-1})\,.
\label{amp3}
\end{multline}
It is worth emphasizing that this change of variables is a gauge fixing procedure, whose goal is to put to the identity one of the $h_s$ (here $h_s =h_1$) \emph{in the internal faces}.  Indeed, the diagonal invariance of the field \eqref{inv} induces an invariance under an action of $G$ at each node of each graph. Like in \cite{freidel-louapre-PR1}, this gauge invariance can be (partially) fixed by contracting a maximal tree in the graph. The new point here is that this procedure has to be implemented on graphs with external legs. As it can be observed in the above example, the net result is that $h_1$ indeed disappears from the internal faces, but then appears in the three \emph{external} faces. We will generalize this procedure to any graph of the 2-point function in the next paragraph.

The reader may worry that the symmetry between the three strands is not explicit anymore after this change of variables. Still the symmetry will be completely explicit in our final result. 

Using the Gaussian approximation \eqref{asymp heat}, one gets,
\beq \label{amp5}
A_{\mathcal G; t}(\{g_s\};\{\tilde{g}_s\}) \simeq \Lambda_t^9\,\int dk_2 dk_3\
C(\{g_1,g_2,g_3\}; \{\tilde{g}_1,\tilde{g}_2\,k_2,\tilde{g}_3\,k_3\})\
e^{-\frac{1}{4t}\bigl(|k_2|^2+|k_3|^2+|k_2 k_3^{-1}|^2\bigr)} .
\ee
Here the integral over $h_1$ has been reabsorbed to form 
the propagator \eqref{propa}, with insertion of $k_2, k_3$ in the strands $2$ and $3$.

{\bf The saddle point -} We are now in position to perform a saddle point approximation around:
\beq
k_2\,=\, k_3\,=\,\unit\;.
\ee
The fact that there is a single saddle point can be understood from a larger perspective, as we explain in  Sections \ref{sec:gf}, \ref{sec:simplyconnected}. Mainly, the amplitude of a graph in the Boulatov-Ooguri model is a special case of the flat spin foam model considered in \cite{Bonzom:2010zh}, where it is shown that the integral is generically localized around the set of homomorphisms from the fundamental group of the 2-complex to the group $G$. Here, $\cG$ is simply connected, so there is a single homomorphism to $\SU(2)$.

The saddle point approximation proceeds by expanding $k_2 = e^{X_2}$ into powers of the Lie algebra element $X_2$, and the same for $k_3 = e^{X_3}$. The integrals over $k_2, k_3$ are changed to integrals on the tangent space, the Lie algebra $\su(2)$, equipped with the Lebesgue measure $d^3X_2\, d^3X_3$. The Haar measure is written as:
\beq
dk_2\ dk_3 \,=\, \mu(X_2,X_3)\ d^3X_2\ d^3X_3\;.
\ee

{\bf The Hessian -} Similarly, the function $(|k_2|^2+|k_3|^2+|k_2 k_3^{-1}|^2)$
 in the exponential is expanded to extract the Hessian,
\beq \label{expand distance}
|k_2|^2+|k_3|^2+|k_2 k_3^{-1}|^2 = X_2^2 + X_3^2 
+ \bigl(X_2 - X_3\bigr)^2 + S_{\geq 3}(X_2,X_3)\;.
\ee
The squared quantities in this formula correspond to the norm of the Killing form (in our simple $\SU(2)$ case, this is the Euclidean norm on $\R^3$). We have packed all terms of order 3 and higher into the remainder $S_{\geq 3}(X_2, X_3)$. Quite clearly, the quadratic part is non-degenerate, its kernel being $\{X_2=X_3=0\}$.

In an orthonormal basis for $(X_2^i,X_3^j)$  the Hessian matrix is:
\beq
\Hess = \begin{pmatrix} 2\, I_3 & -I_3 \\ -I_3 & 2\,I_3 \end{pmatrix},
\ee
where $I_3$ is the $3\times 3$ identity matrix. Up to a constant, its inverse is:
\beq \label{invhess}
\Hess\mone = \operatorname{const}\times\ 
\begin{pmatrix} 2\, I_3 & I_3 \\ I_3 & 2\,I_3 \end{pmatrix}.
\ee

{\bf Expanding the propagator -} Since it is a product of Dirac delta, we first consider derivatives of delta. To avoid test functions, we will regard them through the Fourier expansion. Remember that the character $\chi_j$ in the representation of spin $j$ is the trace $\Tr_j$ on the carrier space of dimension $(2j+1)$. This provides us with a matricial picture, in which we can write, at least formally:
\bea
\delta(g\, e^{X}) = \sum_{n =0}^\infty \; \frac{1}{n!}
\sum_{j\in\N/2} (2j+1)\, {\Tr}_j \left(g\, (X)^n \right) \;.
\label{taylor}
\eea
Next, pick up any basis $(J_i)_{i=1,2,3}$ of the Lie algebra (anti-hermitian generators), 
and set\footnote{We will systematically use the sum over repeated indices 
for contractions on the Lie algebra.} $X = X^i\,J_i$, so that
\eqref{taylor} becomes
\bea
\delta(g \, e^{X}) = \sum_{n =0}^\infty  \frac{X^{i_1}X^{i_2}\dotsm X^{i_n}}{n!}
\sum_{j} (2j+1)\, {\text{Tr}}_j \left(g  J_{i_1}J_{i_2}\dots J_{i_n} \right) \;.
\label{taylor2}
\eea
Generators of the Lie algebra form a basis of, say, left invariant vector fields, 
so that the above formula is really a way of computing Lie derivatives 
with matrices (along the vector field $X_{|g} = L_{g*}X$).

Using the above formulae, we can give a meaning to the derivatives of the propagator. 
To keep notations simple at this stage (the details are given in the next sections), 
we write its expansion in a very symbolic form:
\beq
C(\{g_1,g_2,g_3\}; \{\tilde{g}_1,\tilde{g}_2\,e^{X_2},\tilde{g}_3\,e^{X_3}\}) = \sum_{n=0}^{\infty} (X)^n\ C^{(n)}(\{g_1,g_2,g_3\}; \{\tilde{g}_1,\tilde{g}_2,\tilde{g}_3\})\;,
\ee
where $(X)^n$ stands for all products of order $n$ of the form: $\frac{1}{k!\,(n-k)!}X_2^{i_1}\dotsm X_2^{i_k} X_3^{j_1}\dotsm X_3^{j_{n-k}}$, and $C^{(n)}$ the corresponding derivatives evaluated at $X_2=X_3=0$.

Notice that the apparent symmetry breaking between the three strands initially 
introduced appears here as there are no derivatives of the propagator 
with respect to variables of the strand $1$.

{\bf The form of the expansion -} Finally, the localization of the amplitude \eqref{amp5} leads to:
\begin{multline}
A_{\mathcal G; t}(\{g_s\};\{\tilde{g}_s\}) \simeq \Lambda_t^9\,\int d^3X_2\, d^3X_3 \Bigl[\sum_{n=0}^{\infty} \frac{1}{n!}\ (X)^n\ C^{(n)}(\{g_s\}; \{\tilde{g}_s\})\Bigr]\,e^{-\frac{X_2^2 + X_3^2 + (X_2 - X_3)^2}{4t}}\\
\mu(X_2,X_3) e^{-\frac{S_{\geq 3}(X_2, X_3)}{4t}}\;.
\end{multline}
Usually, the saddle point method is applied to evaluate a quantity as an asymptotic series, here in powers of $\Lambda_t$. In that case, at each order $\Lambda_t^k$, we would encounter a lot of terms, mixing the expansions of the measure $\mu$, of the remainder $S_{\geq 3}$, and that of the propagator. However, this is not what we are looking for. Instead, we will organize the series according to the order of the \emph{derivatives of the propagator}. This means that we push the sum over $n$ in the above formula out of the integral, together with the derivatives of the propagator:
\beq
A_{\mathcal G; t}(\{g_s\};\{\tilde{g}_s\}) \simeq \sum_{n=0}^{\infty} C^{(n)}(\{g_1,g_2,g_3\}; \{\tilde{g}_1,\tilde{g}_2,\tilde{g}_3\})\ \rho^{(n)}_t\;,
\ee
with $\rho^{(n)}_t$ being:
\beq \label{def rhon}
\rho^{(n)}_t = \Lambda_t^9\,\int d^3X_2\, d^3X_3 \ (X)^n\ e^{-\frac{X_2^2 + X_3^2 + (X_2 - X_3)^2}{4t}}
\mu(X_2,X_3) e^{-\frac{S_{\geq 3}(X_2, X_3)}{4t}}\;.
\ee
It is \emph{this} coefficient that we want to expand into powers of $\Lambda_t$. 
Actually, our goal is more humble than that, since we just want 
to know whether the leading order of $\rho^{(n)}_t$ converges or diverges with $\Lambda_t$ !

The components of $\rho^{(n)}_t$ satisfy some rotation invariance that we now describe. Let $\mathcal{H}_1$ denote the Lie algebra $\su(2)$ seen as the vector space $\R^3$ equipped with the standard vector representation of $\SU(2)$. Let us also make explicit the vector indices of $\rho^{(n)}_t$:
\beq
\rho^{(n)\,i_1\dotsc i_n}_t = \Lambda_t^9\,\int d^3X_2\, d^3X_3 \ X^{i_1}\dotsm X^{i_n}\ e^{-\frac{X_2^2 + X_3^2 + (X_2 - X_3)^2}{4t}}
\mu(X_2,X_3) e^{-\frac{S_{\geq 3}(X_2, X_3)}{4t}}\;,
\ee
where each insertion $X^i$ is either $X_2^i$ or $X_3^i$. It turns out that $\rho^{(n)\,i_1\dotsc i_n}_t$ is a invariant tensor in $\mathcal{H}_1^{\otimes n}$, that is an intertwiner from $\mathcal{H}_1^{\otimes n}$ to $\C$.

\begin{lemma} \label{lem:rotinv}
 Let $g\in\SU(2)$ and $R(g)$ denote its matrix in the vector representation. Then,
 \beq
 R^{i_1}_{\phantom{i_1}j_1}(g)\,\dotsm R^{i_n}_{\phantom{i_n}j_n}(g)\ \rho^{(n)j_1\dotsc j_n}_t = \rho^{(n)\,i_1\dotsc i_n}_t\;.
 \ee
\end{lemma}

The proof is straightforward. First, we know that the Haar measure $dk_2 dk_3$ on $\SU(2)^2$ is invariant under the conjugation of $k_2$ and $k_3$ by $g$. This implies that the measure $d^3X_2 d^3X_3\,\mu(X_2, X_3)$ is invariant under the rotation by $R(g)$. Then, we apply the same reasoning with the small $t$ heat kernel approximation. The distances on the group: $\vert k_2\vert, \vert k_3\vert$ and $\vert k_2 k_3\mone\vert$ are invariant under conjugation by $g$, and thus their expansion around the saddle point (the right hand side of \eqref{expand distance}) is invariant under the rotation by $R(g)$. Thus, a change of variables $(X_2, X_3)\mapsto (R(g) X_2, R(g)X_3)$ in the above definition of $\rho^{(n)\,i_1\dotsc i_n}_t$ leads to the desired result.

\bigskip

{\bf The zeroth order: mass renormalization -} The leading term to 
the zeroth order $n=0$ is simply obtained by setting $\mu\, e^{-S_{\geq 3}/4t}\simeq 1$, so that: 
\beq
\rho^{(0)}_t = \Lambda_t^{9-6}\ K^{(0)} \,+\, o(\Lambda_t^3)\;,
\ee
where the constant $K^{(0)}$ is basically the inverse square root of the determinant of the Hessian, and is independent of $t$. Since the first correction to the propagator coming from the graph of Figure \ref{fig:2point} is a divergent factor $\Lambda_t^3$ times the bare propagator, it gives the mass renormalization:
\beq
A_{\mathcal G; t}(\{g_s\};\{\tilde{g}_s\}) \simeq \Lambda_t^3\,K^{(0)}\ C(\{g_1,g_2,g_3\}; \{\tilde{g}_1,\tilde{g}_2,\tilde{g}_3\})\;.
\ee
Let us go now to the derivatives of the bare propagator !

{\bf The first order is vanishing -} This is a simple corollary of Lemma \ref{lem:rotinv}. There is a single invariant vector in $\mathcal{H}_1$: the zero vector. Hence,
\beq
\rho^{(1)}_t = 0,
\ee
to all orders in $\Lambda_t$.

{\bf Even orders and the need for a wave function renormalization -} Since the even Gaussian moments are non-zero, it is clear that the leading term to $\rho^{(2m)}_t$ corresponds to approximating the measure to $\mu\simeq1$ and neglecting the expansion $S_{\geq 3}$. Further, the $(2m)$-th Gaussian moment picks up an extra factor $\Lambda_t^{-2m}$ compared with the above zeroth order:
\beq
\rho^{(2m)}_t \simeq \Lambda_t^{3-2m}\ \frac{1}{\pi^m}\int d^3X_2\,d^3X_3\ (X)^{2m}\ e^{-(X_2^2 + X_3^2 + (X_2-X_3)^2)}\;.
\ee
Let us first focus on the scaling properties. It turns out that $\Lambda_t^{3-2m}$ goes to zero with $t\rightarrow0$, as soon as $m\geq 2$. However, for $m=1$, the 2-point function receives a divergent contribution, coming with second derivatives of the bare propagator:
\beq \label{div terms}
A_{\mathcal G; t}(\{g_s\};\{\tilde{g}_s\}) \simeq  C(\{g_s\}; \{\tilde{g}_s\})\,\Bigl(\Lambda_t^3\,K^{(0)} + o(\Lambda_t^3)\Bigr) + C^{(2)}(\{g_s\}; \{\tilde{g}_s\})\,\Bigl(\Lambda_t\, K^{(2)} + o(\Lambda_t)\Bigr).
\ee
This is one of the main result of our analysis, and the details of the terms which are contained in $C^{(2)}$ are reported to the following section.

{\bf Odd orders are finite -} Finally, we can show that there are no divergences in front of odd derivatives of the propagator. Since odd Gaussian moments are zero, non-zero contributions to $\rho^{(2m+3)}_t$ involve the expansions of the measure $\mu$ and $S_{\geq 3}$. More precisely, we have to consider odd terms in their expansion, to create an even term with $(X)^{2m+3}$. A key observation is that the expansion of the measure does not contain linear terms in $X_2, X_3$,
\beq
\mu(X_2, X_3) = 1 + o(X_2) + o(X_3)\;.
\ee
Then, the expansion of $S_{\geq 3}$ begins like:
\beq
e^{-\frac{S_{\geq 3}(X_2, X_3)}{4t}} = 1 - \frac{1}{4t}\,\sum_{s, u,v = 2,3} S^{(3)suv}_{ijk}\,X_s^i\,X_u^j\,X_v^k
 + o(X^3)\;.
\ee
Thus, the first non-zero terms in $\rho^{(3)}_t$ comes from the contraction of $S^{(3)suv}_{ijk}$ with Gaussian moments of order $6$. The latter scale like $\Lambda_t^{-3}$, so that the divergences are fully exhausted through \eqref{div terms}.

\subsection{The second derivatives of the propagator}

Since the first radiative correction exhibits a divergence which cannot be reabsorbed into the bare parameters of the action, the corresponding terms will need to be added to the action from the beginning 
(see the discussion in Section \ref{concl}). So it is worth giving some details on these terms which involve the second derivatives of the propagator. In particular, it is not clear from our saddle point analysis why the corrections should be symmetric in the \emph{three} strands, and not only the strands $2,3$.

{\bf Symmetries of the Gaussian moments -} The coefficient $\rho^{(2)}_t$ in front of the second derivatives of $C$ contains three types of terms:
\begin{align} \label{rho2}
\rho^{(2)ij}_{t,ss} &= \Lambda_t\ \frac{1}{2\pi}\int d^3X_2\,d^3X_3\ X_s^i\,X_s^j \ e^{-(X_2^2 + X_3^2 + (X_2-X_3)^2)}\;, \quad \text{for}\ s=2,3,\\
\rho^{(2)ij}_{t,23} &= \Lambda_t\ \frac{1}{\pi}\int d^3X_2\,d^3X_3\ X_2^i\,X_3^j \ e^{-(X_2^2 + X_3^2 + (X_2-X_3)^2)}\;.
\end{align}
The indices $i,j$ are indices of 3-vectors (thus transforming under the vector representation of $\SU(2)$), and are chosen for convenience to be those of an orthonormal basis (the covariance can be restored using the metric instead of Kronecker delta in the following).

Let $g\in\SU(2)$, and denote $R(g)$ the matrix of the vector representation. Lemma \ref{lem:rotinv} states:
\beq
R(g)^i_{\phantom{i}k}\,R(g)^j_{\phantom{j}l}\ \rho^{(2)kl}_{t,rs} = \rho^{(2)ij}_{t,rs}\;.
\ee
Thus, integrating over all rotations, we see that $\rho^{(2)ij}_t$ is zero unless $i=j$. Furthermore, the three diagonal terms are equal:
\beq
\rho^{(2)ij}_{t,rs} = \delta^{ij}\ \frac{1}{3}\delta_{kl}\,\rho^{(2)kl}_{t,rs}\;.
\ee
Using the symmetry under the exchange of $X_2, X_3$, one also gets $\rho^{(2)ij}_{t,22} = \rho^{(2)ij}_{t,33}$.

{\bf The derivatives of the propagator -} Now let us come to the consequences of these symmetries in the form of the radiative corrections. More precisely, they teach us that one should not consider arbitrary combinations of the second derivatives of the propagator. Indeed, let us focus on the case $r=s=2$ first. Since the Gaussian moments produce a Kronecker $\delta^{ij}$, we consider:
\begin{align}
C^{(2)}_{t;22} &= \int dh\, \delta\bigl(g_1 h \tilde{g}_1\bigr)\,\delta\bigl(g_3 h \tilde{g}_3\bigr)
\sum_{j_2}(2j_2+1) \Tr_{j_2}\bigl(g_2\,h\ J_i\,J_j\, \delta^{ij}\ \tilde{g}_2\bigr)\crcr
&= \int dh\, \delta\bigl(g_1 h \tilde{g}_1\bigr)\,\delta\bigl(g_3 h \tilde{g}_3\bigr) \sum_{j_2}(2j_2+1)\,\bigl[-j_2(j_2+1)\bigr]\, \Tr_{j_2}\bigl(g_2 h \tilde{g}_2\bigr)
\end{align}
in which we have recognized the Casimir $J^2 = -j(j+1)$ in the representation of spin $j$. A similar formula holds for the strand $3$.

We also have to consider crossed terms in $C^{(2)}$ with one derivative on the strand $2$ and one on $3$,
\beq
C^{(2)}_{t;23} = \sum_{j_1,j_2, j_3} (2j_1+1)(2j_2+1)(2j_3+1) \int dh\ \Tr_{j_1}\bigl(g_1 h \tilde{g}_1\bigr)\ \Bigl[\delta^{ij}\ \Tr_{j_2}\bigl(g_2 h\, J_i\,\tilde{g}_2\bigr)\,\Tr_{j_3}\bigl(g_3 h\, J_j\,\tilde{g}_3\bigr)\Bigr].
\ee
It turns out that the integral over $h$ here is crucial. Indeed it projects onto the sector which is invariant under the group action, so that the standard recoupling theory of $\SU(2)$ can be applied \cite{varshalovich-book}. The insertion of $\delta^{ij} J_{i}\otimes J_j$ between the strands $2$ and $3$ can be seen as a grasping in the vector representation (which is obviously the spin 1). This grasping actually has a diagonal action (when the integral over $h$ is really taken into consideration),
\begin{align}
C^{(2)}_{t;23} &= \sum_{j_1,j_2, j_3} (-1)^{j_1+j_2+j_3+1} N_{j_2}N_{j_3}\begin{Bmatrix} j_2 &j_2 & 1\\j_3 & j_3 &j_1\end{Bmatrix}\int dh\, \prod_{s=1,2,3}(2j_s+1) \Tr_{j_s}\bigl(g_s h \tilde{g}_s\bigr) \crcr
&= \sum_{j_1,j_2, j_3} \frac{1}{2}\bigl[j_2(j_2+1) + j_3(j_3+1) - j_1(j_1+1)\bigr]
\int dh\, \prod_{s=1,2,3}(2j_s+1) \Tr_{j_s}\bigl(g_s h \tilde{g}_s\bigr). 
\label{offdiag}
\end{align}
In the first line, we have introduced some coefficients $N_j = \sqrt{(2j+1)j(j+1)}$ and a Wigner 6j-symbol with a spin being 1. The latter is then explicitly evaluated to arrive at the second line.

{\bf The final evaluation -} So the result of this operator is actually a linear combination of Casimirs, which could just be reabsorbed into Laplace operators. But we want to see what combination of Laplace operators comes out, and how the symmetry between the three strands is restored. 

So we cannot really go further just using the symmetries, and we will make use of the explicit Hessian matrix instead. The integral \eqref{rho2} is indeed really standard and produces Gaussian moments given by the matrix elements of the inverse of Hessian \eqref{invhess}. So we contract the derivatives with the Gaussian moments to get:
\begin{align}
&\frac{1}{2!}\bigl(\Hess\mone\bigr)_{22} C^{(2)}_{t;22} + \frac{1}{2!}\bigl(\Hess\mone\bigr)_{33} C^{(2)}_{t;33} + \bigl(\Hess\mone\bigr)_{23} C^{(2)}_{t;23}\\
&\begin{aligned} = \sum_{j_1,j_2,j_3} \Bigl\{-j_2(j_2+1) -j_3(j_3+1) + \frac{1}{2}\bigl[j_2(j_2+1) &+ j_3(j_3+1) - j_1(j_1+1)\bigr]\Bigr\}\\ & \times\int dh\,\prod_{s=1,2,3}(2j_s+1) \Tr_{j_s}\bigl(g_s h \tilde{g}_s\bigr)\end{aligned}\\
&= -\sum_{j_1,j_2,j_3} \bigl[j_1(j_1+1) + j_2(j_2+1) +j_3(j_3+1) \bigr] \ \frac{1}{2}\ \int dh\,\prod_{s=1,2,3}(2j_s+1) \Tr_{j_s}\bigl(g_s h \tilde{g}_s\bigr)\;.
\end{align}
Thus, the explicit form of the Hessian has restored the symmetry between the three strands.

Moreover, the above result can be recast in terms of a differential (Laplace) operator acting on the bare propagator. Indeed, it is known that the characters are eigenfunctions of the Laplace operator $\Delta$ on the 3-sphere, with eigenvalues being the Casimir of the representation,
\beq
\Delta\,\Tr_j(g) = -j(j+1)\, \Tr_j(g)\;.
\ee
Thus, we have our final formula for the expansion of the 2-point graph:
\begin{multline} \label{final 1st corr}
A_{\mathcal G; t}(\{g_s\};\{\tilde{g}_s\}) =  \Bigl(\Lambda_t^3\,K^{(0)} + o(\Lambda_t^3)\Bigr)\ C(\{g_s\}; \{\tilde{g}_s\}) + \Lambda_t\, K^{(2)}\ \left[\sum_{s=1,2,3}\Delta_{(s)}\right]C(\{g_s\}; \{\tilde{g}_s\})\\
+ \text{convergent terms}\;.
\end{multline}

This study of the scaling behavior of the 2-point function is clearly suggestive about which kind of terms have to be added in the Lagrangian at the very beginning, before re-discussing the renormalizability of the 3d model.  A Laplace operator on the group needs be considered in the dynamics in order to achieve a consistent renormalizability of the Boulatov model. Further comments on such a theory are reported in Section \ref{concl}.

We will find quite generally (in any dimensions greater than $3$ see the next section) in Section \ref{sec:simplyconnected} that there are divergences with second derivatives of the propagator, which cannot be reabsorbed with a counter-term in the initial action. But we have not worked out their precise form like the 3d Boulatov case. This is work in progress.

\subsection{$D$-dimensional extension} \label{sect:2}

The similar graph ${\mathcal G}$ (see Fig. \ref{fig:prop2}) in dimension $D$ has the following amplitude:
\bea
A_{\mathcal G}(\{g_s\};\{\tilde{g}_s\}) = 
\int  \prod_{\ell=1}^D dh_\ell  \left[
\prod_{s=1}^D
\delta 
\left(g_{s}
h_s  (\tilde{g}_{s})^{-1}\right) \right]
 \prod_{1 \leq a < b \leq D}
\delta 
\left( h^{\phantom{-1}}_a h^{-1}_b  \right) \;,
\label{amp16}
\eea
where $\{g_s \}_{s=1}^D$ and  $\{\tilde{g}_s\}_{s=1}^D$ are the external group elements.

The computation follows exactly the steps of the previous case. One can reabsorb $h_1$ into each $h_s$, $s=2,\dotsc,D$, so that it disappears from all internal faces, but appears instead in all external faces. The corresponding integral is then seen as the group averaging defining the bare propagator. The regularization proceeds by using the small time behaviour of the heat kernel on all internal faces. Then, one perform a saddle point approximation around $h_s=\unit$, for $s=2,\dotsc,D$. One expands $h_s = e^{X_s}$ into powers of $X_s\in\su(2)$, and the propagator is correspondingly expanded. This yields:
\beq
A_{\mathcal G;t}(\{g_s\};\{\tilde{g}_s\}) = \sum_{n\geq 0} C^{(n)}(\{g_s\};\{\tilde{g}_s\})\ \rho^{(n)}_t,
\ee
with 
\beq
\rho^{(n)}_t = \Lambda_t^{\frac{3}{2}D(D-1)} \int \left[\prod_{s=2}^D d^3X_s\right]\ (X)^n\ \exp\,-\frac{1}{4t}\left(\sum_{s\geq 2} X_s^2 + \sum_{2\leq r<s\leq D}(X_r-X_s)^2\right)\ R(X_s)\;.
\ee
Here $R(X_s)$ is everything which comes from higher order expansions of the Riemannian distances on the manifold and of the Haar measure. Obviously, the leading order of each $\rho^{(n)}_t$ is evaluated with $R(X_s)\simeq 1$.

Again, the behaviour of $\rho^{(n)}_t$ in terms of its vector indices is easy to obtain using the rotation symmetry of the integrand. The result is that it is an intertwiner, from $\mathcal{H}_1^{\otimes n}$ to $\C$. The expansion then produces:
\begin{itemize}
\item At the zeroth order
\beq
A_{\mathcal G;t}(\{g_s\};\{\tilde{g}_s\}) = \Lambda_t^{\frac{3}{2}(D-1)(D-2)} 
C(\{g_s\};\{\tilde{g}_s\})\ K^{(0)} \; + o(\Lambda_t^{\frac{3}{2}(D-1)(D-2)})\;.
   \label{amp19}
\ee
Not so surprisingly, this is finite for $D=1$ or $D=2$.

\item The only rotation invariant vector in $\mathcal{H}_1$ is zero. Hence, $\rho^{(1)}_t =0$, to all orders in $\Lambda_t$.

\item For the second derivatives of the propagator, it is found again that as a tensor in the vector indices, $\rho^{(n)ij}_t$ is proportional to the metric, say $\delta^{ij}$ in an orthonormal basis of $\mathcal{H}_1$. The Gaussian moment brings a factor $\Lambda_t^{-2}$, so that: 
\begin{multline}
A_{\mathcal G;t}(\{g_s\};\{\tilde{g}_s\}) = \Lambda_t^{\frac{3}{2}(D-1)(D-2)} \biggl[K^{(0)}\ C(\{g_s\};\{\tilde{g}_s\}) \\
+  \Lambda_t^{- 2} K^{(2)}
\Bigl(\frac12 \sum_{s=2}^D \Delta_{(s)} + \sum_{2\leq r < s \leq D} 
\partial_{(r)}^i\otimes \partial_{(s)i}\Bigr) C(\{g_s\};\{\tilde{g}_s\}) + o(\Lambda_t^{-2})\biggr] \; .
   \label{amp20}
\end{multline}
The Laplace operators come from the diagonal insertions, $X_s^i X_s^j$ in $\rho^{(2)}_t$, and the crossed derivatives from $X_r^i X_s^j$ with $r\neq s$.
\end{itemize}
Beyond this second order, one can unravel subleading divergences, in front of derivatives of the bare propagator to even and odd orders, as long as the involved Gaussian moments are of order $2k$, with: $\frac32 (D-1)(D-2) - 2k \geq 0$.

Exactly like in the 3d Boulatov model, a divergence with second order derivatives is found, which cannot be balanced with a counter-term of the bare action. Thus, the latter has to be amended to include the necessary quadratic operators from the beginning. However, unlike in the 3d case, we only know the expression \eqref{amp20} for these operators, which is {\em not} explicitly symmetric in the exchange of the $D$ strands. We nevertheless know that it is symmetric, though not in an obvious way. The symmetric expression in 3d was obtained by explicitly computing the action of the graspings $\partial_{(r)}\otimes \partial_{(s)}$, using the rotation invariance of the propagator (see equation \eqref{offdiag}). Preliminary computations in the four-dimensional case exhibits a sum of the $D$ Laplace operators. But it is also possible that in $D\geq 4$, there is some room for other rotation invariant operators.

\section{Arbitrary 2-point graphs} \label{sec:gf}

We would like to generalize the previous method to any graph of the 2-point function, that is with an arbitrary number of internal vertices, and to arbitrary dimensions. The recipe we used can be summarized as follows:
\begin{itemize}
 \item Use a gauge fixing procedure to reduce the number of integrals.
 \item Localize the integral thanks to the (regularized) delta functions.
 \item Evaluate the \emph{external} propagators and their derivatives on the saddle point.
\end{itemize}
Since the amplitude is again of the form of \eqref{amp1}, these three steps can actually be carried out for any graphs of the theory.

However, it will generically be impossible to extract the bare propagator and its derivatives like we did before, for the following reason. The localization, on the second step, takes place on the set $\mathcal{F}$ of flat connections on the 2-complex. This set (see below) is determined by the fundamental group $\pi_1(\cG)$ and is generically an algebraic variety (when $G$ is an algebraic group). 
Exactly like in the weak coupling limit of 2d Yang-Mills, \cite{2d-YM}, one can recast the amplitude as an integral over the normal bundle to $\mathcal{F}$, then perform the integrals over the normal fibers (the transverse directions to $\mathcal{F}$) thanks to the Gaussian behaviour of the heat kernels in these directions, \cite{Bonzom:2010zh}.

The net result is that when applying the third step in the above recipe, the evaluation of the external propagators does not produce the bare propagator of the model, but instead one faces some averaging of it over the set $\mathcal{F}$ of flat connections on $\Gamma$. This set is potentially non-trivial, which means that the group elements on the edges of $\Gamma$ to be integrated \emph{cannot} be localized on the identity. This is obviously in contrast to what happens in the previous section with the presence of a \emph{single} saddle point, $k_2=k_3=\unit$.

Thus, to get the same kind of expansion as that of the previous section, we will later restrict attention to graphs with a trivial internal fundamental group, that we call \emph{simply connected 2-point graphs}.

\subsection{The gauge-fixing procedure}

To understand the gauge fixing procedure, it is useful to remember how the amplitude \eqref{amp1} is built as the partition function (with external data) for a system of connections on a 2-complex. Let $L(\mathcal{G})$ be the set of internal lines (edges). A discrete connection $A$ is a map from $L(\cG)$ to $G$, or equivalently, a collection of group elements assigned to the lines,
\beq
A=(h_l)_{l\in L(\mathcal{G})}\;.
\ee
These group elements can be thought of as parallel transport operators, or holonomies, along each line, between their source and target vertices. The curvature is expected to describe the effect of parallel transport along a closed path in $\cG$. So if $F^{\inte}(\mathcal{G})$ is the set of internal faces, we consider the oriented product of the holonomies along the boundary of each face. This defines a map:
\beq
H \,:\, A \mapsto \bigl( H_f 
= \prod_{l \in \partial f} h^{\epsilon_{lf}}_l\bigr)_{f\in F^{\inte}(\cG)}\;,
\ee
where ${\epsilon_{lf}}=\pm 1$ depends on the relative orientation of the line $l$ and the face $f$.

We assume that the three open, external faces are each identified by a strand $s=1,2,3$, which goes from one external leg to the other\footnote{This is {\em the} hypothesis of our paper which is not generic in ordinary group field theory. Still, it does hold without restriction in the colored model \cite{gurau}.}. The parallel transport operators along them are inserted between the boundary variables, say $g_s$ on the left and $\tilde{g}_s$ on the right,
\beq
H_s(g_s,\tilde{g}_s, A) 
= g_s\, \Bigl[\prod_{l \in \partial f_s} h^{\epsilon_{lf}}_l\Bigr]\, (\tilde{g}_s)^{-1}\;.
\ee
The amplitude then reads:
\beq
A_\cG(\{g_s\};\{\tilde{g}_s\}) = \int \prod_{l\in L(\cG)} dh_l\ \prod_{f\in F^{\inte}(\cG)} \delta\Bigl( H_f(A)\Bigr)\ \prod_{s=1,2,3} \delta\Bigl(H_s(g_s,\tilde{g}_s, A)\Bigr)\;.
\ee
Like in the previous section, we regularize the internal delta functions with heat kernels that are approximated by their small time behaviour:
\beq \label{reg amp}
A_{\cG;t}(\{g_s\};\{\tilde{g}_s\}) = \Lambda_t^{(\dim G)\vert F^{\inte}\vert} \int \prod_{l\in L(\cG)} dh_l\ \prod_{f\in F^{\inte}(\cG)} e^{-\frac{ \vert H_f(A)\vert^2}{4t}}\ \prod_{s=1,2,3} \delta\Bigl(H_s(g_s,\tilde{g}_s, A)\Bigr)\;,
\ee
where $F^{\inte}$ is the number of internal faces.

We now come to the gauge fixing. It seems at first that the standard procedure is not available due to the external legs, since there are boundary variables $(g_s, \tilde{g}_s)$ which are not integrated. However the integrals over the group elements $h_l$ on the links adjacent to the external vertices are sufficient to get the standard gauge invariance. Let us say:
\begin{lemma}
 Inserting and integrating over some parallel transport group elements $h, \tilde{h}$ on the external legs of any 2-point graph leaves the amplitude invariant:
 \beq \label{restore sym}
 \int_{G^2} dh\,d\tilde{h}\ A_{\cG;t}(\{g_s\,h\};\{\tilde{g}_s\,\tilde{h}\}) = A_{\cG;t}(\{g_s\};\{\tilde{g}_s\})\;.
 \ee
 (The result actually holds even without integrating).
\end{lemma}

The proof is straightforward. Consider one of the external vertices, say that on the left $v^{\ext}$. Assume without loss of generality that the three internal adjacent links to $v^{\ext}$ are oriented outwards. Let $h\in G$. From a connection $A$, define a new connection ${^{(h)}}A$ by changing the group elements on the three adjacent links to $v^{\ext}$ to $k_l = h_l\,h\mone$. Let us investigate how it modifies \eqref{reg amp}. This change of variables leaves the measure invariant: $\prod_l dh_l = \prod_l dk_l$. Quite clearly also, the curvature around the internal faces is unchanged: $H_f(A) = H_f({^{(h)}}A)$. However, it does change the holonomies along the external faces to:
\beq
H_s(g_s,\tilde{g}_s, {^{(h)}}A) = H_s (g_s\,h, \tilde{g}_s,A)\;.
\ee
Since the amplitude is all in all independent of $h$, we can integrate it as well, with the normalized Haar measure ($G$ is compact). Then, repeat the process on the second external vertex $\tilde{v}^{\ext}$.

What we have gained is that the integrand is now explicitly invariant under a natural group action acting on all the vertices (and not only internal). This is an action of $G^{\vert V(\cG)\vert}$ parametrized by a collection of group elements attached to the vertices, $g = (g_v)$, which changes a connection $A$ as well as $h, \tilde{h}$ to:
\beq
{^{(g)}}A = \bigl(g_{t(l)}\,h_l\,g_{s(l)}\mone\bigr)_{l\in L(\cG)},\quad {^{(g)}}h = h\,g_{v^{\ext}}\mone,\quad {^{(g)}}\tilde{h} = \tilde{h}\,g_{\tilde{v}^{\ext}}\mone\;.
\ee
This symmetry can be used to gauge-fixed the integrand through a standard procedure, \cite{freidel-louapre-PR1}, which amounts to setting $h_l=\unit$ on every line of a maximal tree $\mathcal{T}$ on $\cG$. The tree touches all vertices of $\cG$, without forming a loop, and does not contain the external lines. Notice that the introduction of the variables $h, \tilde{h}$ in \eqref{restore sym} is a matter of convenience. That makes it possible to apply the result of \cite{freidel-louapre-PR1} as it is. Without introducing the additional variables $h,\tilde{h}$, the same result could have been achieved, but with a less straightforward proof, in which what happens on the external vertices $v^{\ext}, \tilde{v}^{\ext}$ should have been carefully studied.

Let us write $\cG^{(\mathcal{T})}$ the deformation retract of $\cG$ along the tree $\mathcal{T}$. Thus $\cG^{(\mathcal{T})}$ is a 2-complex with a \emph{single} vertex, with $(\vert L(\cG)\vert - \vert V(\cG)\vert +1)$ lines, while the number of faces is unchanged. A connection $A$ is redefined to be collection of group elements on $\cG^{(\mathcal{T})}$. The result of the gauge-fixing process leads to:
\begin{multline}
A_{\mathcal G; t}(\{g_s\};\{\tilde{g}_s\}) = \Lambda_t^{(\dim G)\vert F^{\inte}\vert} 
\int \prod_{ l \in L(\mathcal{G}^{(\mathcal{T})})} dh_l\ \prod_{f\in F^{\inte}(\cG)} e^{-\frac{ \vert H_f(A)\vert^2}{4t}}\\
\left[\int dh\,d\tilde{h} \prod_{s=1}^3
\delta 
\Bigl(g_{s} h [\prod_{l\in \partial f_s}h_{l}^{\epsilon_{lf}} ] (\tilde{g}_{s} \tilde{h})^{-1}\Bigr)\right].
\end{multline}
It is convenient to get rid of one of the two integrals, say over $h$ in this formula. This is easily achieved by changing the variables according to: $h_l\mapsto h\mone h_l h$, and $\tilde{h}\mapsto \tilde{h}h$. The remaining integral over $\tilde{h}$ is very useful: it can be viewed as the group averaging which defines the bare propagator in \eqref{propa} ! Thus, we reach the following
\begin{lemma} \label{lemma:gf}{Gauge-fixing on a tree of a 2-point graph -} The amplitude of a 2-point graph $\cG$ can be reduced after retraction of a maximal tree to $\cG^{(\mathcal{T})}$, to:
\beq \label{amp after gf}
A_{\mathcal G; t}(\{g_s\};\{\tilde{g}_s\}) = \Lambda_t^{(\dim G)\vert F^{\inte}\vert} 
\int \prod_{ l \in L(\mathcal{G}^{(\mathcal{T})})} dh_l \prod_{f\in F^{\inte}(\cG)} e^{-\frac{ \vert H_f(A)\vert^2}{4t}}
C(\{g_{s}\prod_{l\in \partial f_s}h_{l}^{\epsilon_{lf}} \} ,  \{\tilde{g}_{s}\})\;.
\ee
The map $H$ which sends the elements $h_l$ attached to the remaining edges to group elements on faces gives a presentation of the fundamental group $\pi_1(\cG)$ (see below).
\end{lemma}

\subsection{The generic saddle point analysis}

The equation \eqref{amp after gf} gives us the opportunity to get the leading order of the amplitude for an arbitrary 2-point graph, using the results of \cite{Bonzom:2010zh}. Indeed, the regularized delta, here with a Gaussian behaviour, clearly enforces a localization on the set of \emph{flat connections} $\mathcal{F}$ defined as:
\beq
\mathcal{F} = H\mone(\unit)\;.
\ee
This set admits a nice (and well-known) geometric description in terms of the 2-complex $\cG^{(\mathcal{T})}$. Indeed, one can read a presentation of the fundamental group $\pi_1(\cG)$ from it: there is one generator per edge of $\cG^{(\mathcal{T})}$, and one relation per face. The generators are in one-to-one correspondence with the group elements to be integrated in \eqref{amp after gf}, and the relations among the generators are exactly the conditions enforced by the (Gaussian regularized) Dirac delta in the amplitude (see details in \cite{Bonzom:2010zh}). So as expected (that was the initial motivation of Boulatov \cite{boul}), the amplitude is localized around the set $\mathcal{F}$ which is determined by the fundamental group. This is the set of homomorphims of the fundamental group into $G$, also known in the mathematical literature as the \emph{representation variety} of $\pi_1(\Gamma)$ into $G$,
\beq
\mathcal{F} = \Hom\bigl(\pi_1(\Gamma), \SU(2)\bigr)\;.
\ee
Exactly like in the weak coupling limit of 2d Yang-Mills, \cite{2d-YM}, one can then recast the amplitude as an integral over the normal bundle to $\mathcal{F}$, then perform the integrals over the normal fibers (the transverse directions to $\mathcal{F}$) thanks to the Gaussian behaviour of the heat kernels in these directions, \cite{Bonzom:2010zh}.

Thus, $\vert H_f(A)\vert^2$ can be linearized in the directions which are transverse to $\mathcal{F}$, around each flat connection $\Phi=(\phi_l)_{l\in L(\cG^{(\mathcal{T})})}$. Let us denote $X_\Phi$ the variations along these directions. Then, the Hessian is given by the quadratic terms in $X_\Phi$:
\beq \label{hessian}
\sum_f \vert H_f(A)\vert^2 = \Vert dH_\Phi(X_\Phi)\Vert^2 + S_{\geq 3;\Phi}(X_\Phi)\;.
\ee
Here $\Vert\cdot\Vert$ is the norm coming from the Killing form on $\alg^{F^{\inte}}$.

To extract the leading order, it is sufficient to evaluate the inserted propagator on $\mathcal{F}$. So the integrals over the normal fibers over each flat connection $\Phi$ can be carried out exactly like in \cite{Bonzom:2010zh}. This gives:
\beq
A_{\mathcal G; t}(\{g_s\};\{\tilde{g}_s\}) \simeq \Lambda_t^{\Omega(\cG)} 
\int_{\mathcal{F}} \vol_{\mathcal{F}}(\Phi)\ C(\{g_{s}\prod_{l\in \partial f_s}\phi_{l}^{\epsilon_{lf}} \} ,  \{\tilde{g}_{s}\})\;,
\ee
where the divergence degree\footnote{This formula is an integral over the non-singular subset of $\mathcal{F}$, on which $\dim(\ker dH_\Phi)^\perp$ is constant.} $\Omega(\mathcal{G})$ is:
\beq
\Omega(\cG) = (\dim G)\vert F^{\inte}\vert - \dim(\ker dH_\Phi)^\perp\;.
\ee
The volume form $\vol_{\mathcal{F}}(\Phi)$ can be expressed using the Reidemeister torsion of the 2-complex $\cG$. Generically, while this formula indeed factorizes the leading divergence, it also shows that the bare propagator \emph{cannot} be extracted. The latter is instead somehow averaged over $\mathcal{F}$.

For instance, if the fundamental group of $\Gamma$ is that of a lens space, $\pi_1(\Gamma) = \Z_p$, for $p\in\N$, then there are non-trivial (reducible but not central) flat connections, and the moduli space, $\mathcal{F}/G$, has a finite number of points. The volume of the orbits of the group action can be factored out, so that one has:
\beq
A_{\mathcal G; t}(\{g_s\};\{\tilde{g}_s\}) \simeq \Lambda_t^{\Omega(\cG)} \sum_{[\Phi]\in\mathcal{F}/G} \nu([\Phi])\ C(\{g_{s}\prod_{l\in \partial f_s}[\phi]_{l}^{\epsilon_{lf}} \} ,  \{\tilde{g}_{s}\})\;.
\ee
While these graph amplitudes do not admit a direct expansion of the kind \eqref{typical exp}, it may be still possible to extract some mass and wave-function renormalization using a prescription and a scale around which the expansion is enforced, like one usually does in ordinary scalar field theory\footnote{In ordinary field theory, a 2-point graph does not take the simple form \eqref{typical exp}. Nevertheless, the mass renormalization $\delta_m$ is typically extracted by evaluating the amplitude at a given external scale $\mu$, and the wave-function counter-term $\delta_Z$ by evaluating the derivatives of the amplitude with respect to the external momenta, at the same scale $\mu$.}. This certainly deserves to be further investigated. 

\section{Simply connected 2-point graphs} \label{sec:simplyconnected}

We will not go further into the study of generic 2-point graphs, and instead focus on a nice class of graphs for which it is possible to extract the bare propagator at the leading order, and its derivatives at the sub-leading orders: the simply connected 2-point graphs. So from now on, we assume that $\cG^{(\mathcal{T})}$ gives a presentation of the trivial group,
\beq
\pi_1\bigl(\cG^{(\mathcal{T})}\bigr) = \langle (a_l)_{l\in L(\cG^{(\mathcal{T})})} ; \bigl(r_f = \prod_{l\in \partial f} a_{l}^{\epsilon_{lf}}\bigr)_{f\in F^{\inte}(\cG)} \rangle = \{\unit\}\;.
\ee
Thus, there is a single flat connection, namely the trivial one,
\beq
\forall l\in L(\cG^{(\mathcal{T})}) \qquad h_l\,=\,\unit\;.
\ee
Combining this fact with the expression \eqref{amp after gf} of the amplitude allows to apply the specific result of \cite{Bonzom:2010ar} concerning simply connected graphs. We briefly outline the reasoning. We will perform a saddle point approximation of \eqref{amp after gf}. The Hessian is generically given in \eqref{hessian}. But on the trivial connection, the differential of the curvature map $H$ simplifies, and is actually the first coboundary operator of the 2-complex $\cG^{(\mathcal{T})}$ (with the external edges and faces removed):
\beq
dH_{\unit} = \delta^1\bigl(\cG^{(\mathcal{T})}, \R\bigr)\;.
\ee
The non-degeneracy of the Hessian is obtained through:
\beq
\ker dH_\unit = H^1\bigl(\cG^{(\mathcal{T})}, \R\bigr) = \{ 0 \}\;,
\ee
where the first equality follows from the fact that $\cG^{(\mathcal{T})}$ has a single vertex, and the second from the Hurewicz theorem.

Then, the following steps are exactly those we performed on the simple graph in Section \ref{sec:ex saddle}. The group elements are expanded into powers of Lie algebra elements, $h_l = e^{X_l}$, and the measure becomes:
\beq
\prod_{l\in L(\cG^{(\mathcal{T})})}dh_l = \prod_{l\in L(\cG^{(\mathcal{T})})} d^3X_l\ \mu(\{X_l\})\;.
\ee
We again expand symbolically the propagator:
\beq
C(\{g_s \prod_{l\in \partial f_s}e^{\epsilon_{lf}X_l}\}; \{\tilde{g}_s\}) = \sum_{n=0}^{\infty} \frac{1}{n!}\ (X)^n\ C^{(n)}(\{g_1,g_2,g_3\}; \{\tilde{g}_1,\tilde{g}_2,\tilde{g}_3\})\;,
\ee
where $(X)^n$ stands for all products of the form: $\prod_l \prod_{k_l=1}^{\beta_l} X_l^{i_{k_l}}$, with $\sum_l \beta_l = n$, $C^{(n)}$ the corresponding derivatives, evaluated at $X_l=0$.

So the amplitude takes the form of an expansion into derivatives of the bare propagator, as we wanted:
\beq
A_{\mathcal G; t}(\{g_s\};\{\tilde{g}_s\}) \simeq \sum_{n=0}^{\infty} \frac{1}{n!} C^{(n)}(\{g_1,g_2,g_3\}; \{\tilde{g}_1,\tilde{g}_2,\tilde{g}_3\})\ \rho^{(n)}_t\;,
\ee
with $\rho^{(n)}_t$ being:
\beq
\rho^{(n)}_t = \Lambda_t^{(\dim G)\vert F^{\inte}\vert}\,\int \prod_l d^3X_l \ (X)^n\ e^{-\frac{\Vert dH_\unit(X_l)\Vert^2}{4t}}
\mu(\{X_l\}) e^{-\frac{S_{\geq 3}(\{X_l\})}{4t}}\;.
\ee

\begin{itemize}
\item Using the same method as that of Lemma \ref{lem:rotinv} (taking advantage of the rotation symmetry of the integrand), it is found that $\rho^{(n)}_t$ seen as a tensor in $\mathcal{H}_1^{\otimes n}$ (in its vector indices) is actually an {\em invariant tensor}.
\item The divergence degree at the zeroth order is:
\begin{align}
\Omega(\cG) &= \bigl(\dim G\bigr)\ \Bigl(\vert F^{\inte}(\cG) \vert - \vert L(\cG)\vert + \vert V(\cG)\vert -1\Bigr)\\
&= \bigl(\dim G\bigr) \Bigl( \chi(\Gamma) - 1\Bigr)\;.
\end{align}
In particular, if $\Gamma$ is the 2-skeleton of a cell decomposition of a closed orientable 3-manifold, then its Euler characteristic is the number of 3-cells of the cell decomposition, and we recover some standard result of the field. The amplitude is thus:
\beq
A_{\mathcal G; t}(\{g_s\};\{\tilde{g}_s\}) = \Lambda_t^{\Omega(\cG)} K^{(0)}\ C(\{g_{s}\} ,  \{\tilde{g}_{s}\}) +o(\Lambda_t^{\Omega(\cG)})\;,
\ee
where the constant $K^{(0)}$ comes from the Gaussian integral.
\item There is no term with first derivatives of the propagator, since $\rho^{(1)}_t =0$ to all orders (because the only invariant vector in $\mathcal{H}_1$ is the zero vector).
\item Next we look at the leading order of $\rho^{(2m)}_t$. It is obtained by evaluating $\mu\,e^{-S_{\geq 3}/4t}\simeq 1$ on the saddle point. The Gaussian moment of order $2m$ brings a factor $\Lambda_t^{-2m}$, so that $\rho_t^{(2m)}$ behaves like:
\beq
\rho_t^{(2m)} \simeq \Lambda_t^{\Omega(\cG)-2m} K^{(2m)}\;.
\ee
Since, the divergence degree $\Omega(\cG)$ can be arbitrarily high, it also means we can have an arbitrarily high number of derivatives of the propagator with divergences. More precisely, this happens for all $m$ such that:
\beq
2m \leq \Omega(\cG)\;.
\ee
\end{itemize}

\section{Towards a new scale analysis of the model}
\label{concl}

We have studied graphs of the 2-point function in the $D$-dimensional extension of the Boulatov model. In the case of a trivial fundamental group (that we call simply connected graphs), we have been able to recast the amplitude as an expansion whose zeroth order term is the bare propagator, and higher orders are given by its derivatives. We have been mainly interested in finding which derivatives come with a divergence (a positive exponent of the ultraviolet cut-off $\Lambda$). Like in identically distributed matrix models, we found that arbitrarily high orders of derivatives of the propagator receive divergent factors, as the considered graphs become larger.

For the first radiative correction to the propagator in $D=3$, at the second order in the coupling constant, we have found two divergences: one renormalizing the mass, and one in factor of second derivatives of the propagator (and derivatives of higher orders in $D\geq 4$). In ordinary field theory (say $\phi^4_4$), the latter renormalizes the operator $\phi \partial^2 \phi$, and is usually reabsorbed by a renormalization of the wave-function. But since the initial action (and hence the propagator) does not contain derivatives of the field, the divergence we observe cannot be reabsorbed as a counter-term from the action. This shows that a renormalizable Boulatov model requires to include such terms in the bare action.

So to make sense of the model, one could either consider the large $N$ limit proposed in \cite{Gurau:2010ba}, or consider a new action in which the quadratic part in the field is of the form:

\beq
S_{kin}[\phi] = \int [\prod_{i=1}^3\,dg_i]\;
\phi(g_1,g_2,g_3) \left(\sum_{s=1}^3 \Delta_{(s)} + m^2\right)
\phi(g_1,g_2,g_3)\;,
\ee
As shown from the computation \eqref{offdiag}, terms which contract a derivative on $g_1$ with another on $g_2$ can be reabsorbed into the Laplace operators. Further, it seems natural to choose the same coupling for the three strands of the propagator, as it actually happens in the first radiative correction, \eqref{final 1st corr}. Such a theory has already been considered in the literature, by Di Mare and Oriti \cite{DiMare:2010zp} (and more references therein) for instance. Classical solutions, with interesting perturbation properties and some effective dynamics have been investigated there. This leads to a scenario consistent with the emergence of matter fields as a phase in the context. Thus, besides the fact that this is one of the most natural action, and other interesting features as we have just mentioned, the term we propose to add appears in the present study as an important quantity concerning the renormalization.

Still, a detailed treatment as a quantum field theory, starting with power-counting results like those obtained in the (standard) Boulatov model, are missing. Let us sketch very basically the analysis of the propagator. After Fourier transform on $\SU(2)^3$, the propagator is given by:
\beq
\tilde{C}(j_1,j_2,j_3) = \frac{1}{\sum_{s=1}^3 j_s(j_s+1) + m^2}\;.
\ee
Back in direct space, and after composition with the projector on the sector invariant under the diagonal group action, the propagator reads:
\beq
C(\{g_s\}, \{\tilde{g}_s\}) 
= \sum_{j_1,j_2,j_3}  \int dh \prod_{s=1}^3 \left[(2j_s+1)\, 
\chi_{j_s} (g_s h \tilde{g}^{-1}_s) \right] \tilde{C}(j_1,j_2,j_3) \;.
\ee

To prepare a scale analysis, we introduce the Schwinger parametric representation of the propagator
\begin{align}
C(\{g_s\},\{\tilde{g}_s\}) &= 
\sum_{j_1, j_2, j_3}  \int dh \prod_{s=1}^3 \left[(2j_s+1)\, 
\chi_{j_s}(g_sh\tilde{g}^{-1}_s) \right] \,
\int_0^\infty dt\, e^{- t\left(\sum_s j_s(j_s+1) +  m^2\right)}\\
&= \int_0^\infty dt \, e^{- tm^2} \int dh\, 
 \prod_{s=1}^3 K_{t} (g_s h \tilde{g}^{-1}_s)\;.
\label{new propa}
\end{align}
In the second line, we have noticed that the sum over the spin of each strand produces exactly the heat kernel on $\SU(2)$. Note also that the latter does not appear as a regularization anymore.

Taking $M>1$, we can see that the form of the propagator makes the introduction of scales very natural, \cite{vincent}. We write $C=\sum_i C_i$, where $C_i$ is the sliced propagator (for the massless situation),
\beq
 C_i(\{g_s\},\{\tilde{g}_s\}) = \int^{M^{-2i}}_{M^{-2(i+1)}} \, dt\, \int dh \prod_{s=1}^3  
K_t (g_s h \tilde{g}^{-1}_s)\;.
\ee
From what we know about the behaviour of the heat kernel, it is clear that at high scales, $i\gg 1$, the time $t$ is small, and hence the propagator non-zero only on a small neighborhood of the set $(g_s h \tilde{g}^{-1}_s)=\unit$. Taking inspiration in the Euclidean case, we may conjecture the following bound for the slice $i$:
\beq
C_i (\{g_s\},\{\tilde{g}_s\}) \leq A\  M^{i(3\dim G\,-2)} \int dh\ \prod_{s=1,2,3} e^{- \delta\, M^{2i}\, \vert g_s h \tilde{g}_s^{-1}\vert^2}\;,
\ee
where $A, \delta$ are some constants. The factor $M^{i(3\dim G\,-2)}$ is the ultraviolet cut-off, while the factor $M^{2i}$ in the exponential rather plays the role of an infrared regulator. Again, at large scales $i$, the propagator will grow but only if $(g_s h \tilde{g}_s^{-1})$ comes closer to the identity.

We hope the rather sketchy picture we have just drawn can be further extended to the amplitude of the graphs themselves, and that a new relevant ``locality principle'' can be reached which would enable renormalization.

Finally we note that it is not clear but certainly very interesting to investigate the non simply connected graphs. Though they do not directly admit a simple expansion like \eqref{typical exp}, it may be possible to extract some mass and wave-function renormalization using a prescription and a external scale around which the expansion can be settled, like in ordinary scalar field theory. Mathematically, we know that the insertions of (closed) knots and links in this theory produces knot polynomials, so it may also be fruitful to consider insertions of open strands like those of the 2-point function.

\section*{Acknowledgements}

The authors are pleased to thank Vincent Rivasseau and Razvan Gurau for fruitful discussions.
Research at Perimeter Institute is supported by the Government of Canada through Industry 
Canada and by the Province of Ontario through the Ministry of Research and Innovation.

\end{document}